\documentstyle[preprint,aps]{revtex}
\begin{document}
\draft
\title{How to generate spinor representations in any dimension in terms of projection operators
}
\author{ N. Manko\v c Bor\v stnik}
\address{ Department of Physics, University of
Ljubljana, Jadranska 19, 1111, \\
and Primorska Institute for Natural Sciences and Technology,\\
C. Mare\v zganskega upora 2, Koper 6000, Slovenia}
\author{ H. B. Nielsen}
\address{Department of Physics, Niels Bohr Institute,
Blegdamsvej 17,\\
Copenhagen, DK-2100
 }

\date{\today}

\maketitle

\begin{abstract} 
We present a method to find solutions of the Weyl or the Dirac equation {\em without 
specifying a representation choice for $\gamma^a$'s}. By taking $\gamma^a$'s
formally as independent variables, we construct solutions out of $2^d$ orthonormal polynomials of 
$\gamma^a$'s (in $d$-dimensional space)
operating on a ``vacuum state''. Polynomials reduce into $2^{d/2}$ 
or $2^{(d+1)/2}$ repetitions of the Dirac spinors for $d$ even or odd, respectively. We further propose
the  corresponding  graphic presentation of basic states, which offers an easy way to see all the quantum 
numbers of states with respect to the generators of the Lorentz group,
as well as transformation properties of the states under any operator. 
\end{abstract}

\newpage
\section{Introduction}
\label{introduction}

If one wants to work with spinor states one a priori has the problem that one must choose a 
certain representation of the $\gamma^a$-matrices before writing down in a safe way any state 
of the spinor. It is therefore often attractive to avoid as long as possible writing down 
explicit spinor states. It is the purpose of the present note  to propose a technique that allows 
us in practice to {\em avoid the open choice of representations }for some time. The technique enables 
accordingly to find solutions of the Weyl or the Dirac equation without making a choice of a special
representation for  $\gamma^a$ operators. The technique is built on 
the property of spinors that the generators of the Lorentz transformations  can be expressed
as binomials of the operators $\gamma^a$, which fulfil the Clifford algebra. Taking these operators
formally as independent variables, we construct $2^d$ orhonormal vectors. Spinor states are then represented in terms 
of polynomials of operators $\gamma^a$, operating on some ``vacuum state''.

In reality we  of course find at the end that the method really turns out to mean that we have 
chosen a special 
representation, although we have retained the notation so that we have continuously stated which one we 
used; thus {\em the usual problem of different articles using different notations could really be avoided.}

The proposed technique was initiated and developed by one of the authors of this note, when  proposing an  
approach\cite{norma92,norma93,normaixtapa01} in which all the internal degrees of freedom of either 
spinors or vectors can be described in the space of $d$-anticommuting (Grassmann) coordinates, if the 
dimension of ordinary space is also $d$. In this approach (two kinds of) $\gamma^a$ operators were defined
which, when operating on a vacuum state, define basic vectors, which are polynomials of anticommuting coordinates.
These $2^d$ polynomials  are orthonormal with respect to the inner product defined as an integral over anticommuting 
coordinates, with  an appropriately chosen weight function. 
Both authors of this note then used the results of this approach in the space of differential forms to    
generalize\cite{holgernormadk}  the approach of 
K\" ahler\cite{kahler64}, who defined in the space of differential forms  spins of spinors. This generalization
also manifests  the property of the $\gamma^a$ operators that may formally be used to define orthonormal states of
spinors.

We demonstrate in this note a simple and transparent technique for 
explicitly writing down, in any dimension $d$ of any signature, spinor representations in terms of  $\gamma^a$ 
operators, for which we only need to know that they fulfil the Clifford algebra and that the generators
of the Lorentz transformations are binomials of $\gamma^a$'s - and not at all how the matrix 
representation of $\gamma^a$'s looks like ( section \ref{clifford}). It really means that we are taking operators 
$\gamma^a$ formally as independent operators. We then choose polynomials of  $\gamma^a$'s to  define, when operating 
on a state, 
which we call $|\psi_0>$, the ``vacuum state'', the orthonormal basis (section \ref{eigenstates}). We  arranged 
the orhonormal basis into irreducible representations of the Lorentz group. Using this technique, it is then 
straightforward, for example, to find  solutions of the Dirac equation, for either massless or massive fermions, 
in any dimension $d$ and for any signature (section \ref{solutions}), or to apply  any operator to a spinor state
(section \ref{irreducible}), or to find irreducible representations of subgroups 
of the Lorentz group $SO(q, d-q),$ where $q $ means the number of time-like coordinates.

We further propose a very simple graphic representation (section \ref{graphic}), which makes the technique transparent 
by enabling us to easily see all the 
quantum numbers of states with respect to the generators of the Lorentz group, as well as how states transform when  
operated on by the operators.

We also define within the same technique what we call ``families'' of irreducible representations 
(section \ref{families}). A ``family'' of polynomials of (formally independent) operators $\gamma^a$,  
when operating 
on a state, represents a Dirac spinor in $d$-dimensional space. States belonging 
to different ``families'' are orthogonal. Usually, of course, we deal with only one ``family'', that is with only 
one Dirac spinor.

We demonstrate (section \ref{demonstration}) the proposed technique as well as
the graphic representation of this technique for $d=3$ and $d=4$ and the Minkowski metric.

In this paper, we  assume an arbitrary
signature of space time so that our metric tensor $\eta^{ab}$, with $a,b \in \{0,1,2,3,5,\cdots \d \}$ 
is diagonal with values $\eta^{aa} = \pm 1$, depending on the chosen signature.

\section{Clifford and Lorentz algebra}
\label{clifford}

In this note we consider spinors only.

Let  operators $\gamma^a$ close the Clifford algebra
\begin{equation}
\{\gamma^a, \gamma^b \}_+ = 2\eta^{ab}, \quad {\rm for} \quad a,b \quad \in \{0,1,2,3,5,\cdots,d \},
\label{clif}
\end{equation}
for any $d$, even or odd, and let the Hermiticity property of $\gamma^a$'s be
\begin{eqnarray}
\gamma^{a+} = \eta^{aa} \gamma^a,
\label{cliffher}
\end{eqnarray}
in order that 
the $\gamma^a$ be unitary as usual, i.e. ${\gamma^a}^{\dagger}\gamma^a=1$.

The inner product of a ket $\gamma^a|\psi>$ and a bra $(\gamma^a|\psi>)^{\dagger}$ is 
a nonnegative real number, so that 
$\gamma^{a+} \gamma^a = I, $ with $I$ the unity operator, if $<\psi|\psi>$ is a nonnegative real number.

 The operators 
\begin{equation}
S^{ab} = \frac{i}{4} [\gamma^a, \gamma^b ] := \frac{i}{4} (\gamma^a \gamma^b - \gamma^b \gamma^a)
\label{sab}
\end{equation}
close the algebra of the Lorentz group 
\begin{equation}
\{S^{ab},S^{cd}\}_- = i (\eta^{ad} S^{bc} + \eta^{bc} S^{ad} - \eta^{ac} S^{bd} - \eta^{bd} S^{ac})
\label{loralg}
\end{equation}
and have the following Hermiticity property
\begin{eqnarray}
S^{ab+} = \eta^{aa} \eta^{bb}S^{ab}.
\label{sabher}
\end{eqnarray}
We also see that these operators of the Lorentz algebra fulfil the  spinor algebra 
\begin{equation}
\{S^{ab},S^{ac}\}_+ = \frac{1}{2} \eta^{aa} \eta^{bc}.
\label{spinalg}
\end{equation}
Recognizing from Eq.(\ref{loralg}) that two operators $S^{ab}, S^{cd}$ with all indices different 
commute, we readily select the Cartan subalgebra of the algebra of the Lorentz group, which has 
a basis of
\begin{eqnarray}
m &=& \quad  d/2, \quad \;\;\; {\rm for} \quad d \quad {\rm even},
\nonumber\\
m &=& (d-1)/2, \quad {\rm for} \quad d \quad {\rm odd},
\label{cartan}
\end{eqnarray}
commuting operators.

It is  useful to define one of the Casimirs of the Lorentz group which determines in even dimensional spaces
the handedness of an irreducible representation of the Lorentz group\footnote{To see the definition of the 
operator $\Gamma$ for any spin in even-dimensional spaces see references\cite{norma93,norma94,bojannorma,%
holgernormadk}.} 
\begin{eqnarray}
\Gamma :&=&(i)^{d/2}\; \;\;\;\;\;\prod_a \quad (\sqrt{\eta^{aa}} \gamma^a), \quad {\rm if } \quad d = 2n, 
\nonumber\\
\Gamma :&=& (i)^{(d-1)/2}\; \prod_a \quad (\sqrt{\eta^{aa}} \gamma^a), \quad {\rm if } \quad d = 2n +1,
\label{hand}
\end{eqnarray}
for any integer $n$. We understand the product of $\gamma^a$'s in the ascending order with respect to 
the index $a$: $\gamma^0 \gamma^1\cdots \gamma^d$. 
Since $(\sqrt{\eta^{aa}} \gamma^a)^{\dagger} =  \sqrt{\eta^{aa}} \gamma^a$ and 
$(\sqrt{\eta^{aa}} \gamma^a)^2 = I$, that is the unit operator, then it follows 
for any choice of the signature $\eta^{aa}$  that 
$\Gamma$ is Hermitean and its square is equal to the unity operator
\begin{eqnarray}
\Gamma^{\dagger}:&=& \Gamma,
\nonumber\\
\Gamma^2 :&=& I.
\label{prophand}
\end{eqnarray}
One also finds that in even-dimensional spaces $\Gamma$ anticommutes while in odd-dimensional spaces
$\Gamma$ commutes with $\gamma^a$'s
\begin{eqnarray}
\{\Gamma,\gamma^a\}_{+} :&=& 0, \quad {\rm for} \quad d \quad {\rm even}, 
\nonumber\\
\{\Gamma,\gamma^a\}_{-} :&=& 0, \quad {\rm for} \quad d \quad {\rm odd}.
\label{prop1hand}
\end{eqnarray}
Accordingly, $\Gamma$ always commutes with the generators of the Lorentz algebra.
In even-dimensional spaces, it is easy to see that eigenstates of the Cartan subalgebra  are  for 
spinors the eigenstates 
of the operator of handedness as well, in odd-dimensional spaces, we choose the basic states to be 
eigenstates of one of the two operators $(1/2)(1 \pm \Gamma)$.

\section{Eigenstates of Cartan subalgebra}
\label{eigenstates}

We shall select operators belonging to the Cartan subalgebra as follows
\begin{eqnarray}
S^{0d}, S^{12}, S^{35}, \cdots, S^{d-2\; d-1}, \quad {\rm if } \quad d &=& 2n,
\nonumber\\
S^{12}, S^{35}, \cdots, S^{d-1 \;d}, \quad {\rm if } \quad d &=& 2n +1.
\label{choicecartan}
\end{eqnarray}
One can easily see that for spinors (Eq.(\ref{spinalg})) the operator of handedness can for 
even-dimensional spaces be written in term of 
the operators of the Cartan subalgebra as follows
\begin{eqnarray}
\Gamma = 2^{d/2} \prod_a \sqrt{\eta^{aa}} \quad S^{0d} S^{12} S^{35} \cdots S^{d-2\; d-1}, \quad {\rm if}
\quad d = 2n.
\label{gammasabeven}
\end{eqnarray}

For odd-dimension we can write
\begin{eqnarray}
\Gamma = 2^{(d-1)/2} \prod_a \sqrt{\eta^{aa}} \quad \gamma^0 
S^{12} S^{35} \cdots S^{d-1\; d}, \quad {\rm if } \quad d = 2n +1.
\label{gammasabodd}
\end{eqnarray}

We can now present  a trivial theorem which  helps, however, to find eigenvectors of the above chosen 
Cartan subalgebra of the Lorentz group.

{\em Theorem 1: } Let $S^{ab}$ be any of the generators of the Lorentz group of Eq.(\ref{sab}). Then
states obtained by operating by the operators
\begin{eqnarray}
\stackrel{ab}{(\pm)}: = \frac{1}{\sqrt{2}}(\gamma^a \pm \sqrt{-\eta^{aa} \eta^{bb}}\; \gamma^b),  
\quad \stackrel{ab}{[\pm]}: = \frac{1}{\sqrt{2}}(1 \pm \sqrt{-\eta^{aa}\eta^{bb}} \;\gamma^a
\gamma^b)
\label{eigensab}
\end{eqnarray}
on any state $|\psi>$, which will not be transformed into zero, are eigenstates of the operator $S^{ab}$
with the eigenvalues $\pm (i/2) \eta^{bb} \; \sqrt{-\eta^{aa} \eta^{bb}}$ and $\pm (i/2) 
(-\eta^{aa}\eta^{bb}) \; \sqrt{-\eta^{aa} \eta^{bb}}$, respectively.

{\em Proof:} To prove this theorem we only have to determine what the  operator $S^{ab} = (i/2)
\gamma^a \gamma^b$, $a\ne b$, 
does when applied to the above states. We find
\begin{eqnarray}
S^{ab} \; (\gamma^a \pm \sqrt{-\eta^{aa} \eta^{bb}}\; \gamma^b)\; |\psi> &=& \pm \frac{i}{2} \; \eta^{bb}\;
\sqrt{-\eta^{aa}\eta^{bb}}\;(\gamma^a \pm \sqrt{-\eta^{aa} \eta^{bb}}\; \gamma^b)\; |\psi>,
\nonumber\\
S^{ab} (1 \pm \sqrt{-\eta^{aa}\eta^{bb}} \;\gamma^a \gamma^b)\; |\psi> &=& \pm \frac{i}{2}\; 
(-\eta^{aa}\eta^{bb}) \sqrt{-\eta^{aa} \eta^{bb}} \; (1 \pm \sqrt{-\eta^{aa}\eta^{bb}} \;
\gamma^a \gamma^b)\; |\psi>,
\label{chequeeigensab}
\end{eqnarray}
which completes the proof. 
According to this theorem, we can construct states which are eigenstates of all the Cartan 
subalgebra operators.

We find that while the operator $(1/2)(1 \pm \sqrt{-\eta^{aa}\eta^{bb}} \;\gamma^a \gamma^b)$ is
the projector
\begin{eqnarray}
(\frac{1}{2} \;(1 \pm \sqrt{-\eta^{aa}\eta^{bb}} \;\gamma^a \gamma^b))^2 = \frac{1}{2}\; 
 (1 \pm \sqrt{-\eta^{aa}\eta^{bb}} \;\gamma^a \gamma^b),
 \label{projector}
\end{eqnarray}
the operator $(\gamma^a \pm \sqrt{-\eta^{aa} \eta^{bb}}\; \gamma^b)$ is a nilpotent operator
\begin{eqnarray}
(\gamma^a \pm \sqrt{-\eta^{aa} \eta^{bb}}\; \gamma^b)^2 = 0.
\label{nilpotent}
\end{eqnarray}

According to the above theorem, it is straightforward to prove the following theorem.

{\em Theorem 2: } For an even dimension ($d=2n$) the states
\begin{eqnarray}
(\gamma^0 \pm \sqrt{-\eta^{00} \eta^{dd}}\; \gamma^d)\;(\gamma^1 \pm \sqrt{-\eta^{11} \eta^{22}}\; \gamma^2)
\cdots (\gamma^{d-2} \pm \sqrt{-\eta^{d-2 d-2} \eta^{d-1 d-1}}\; \gamma^{d-1})|\psi_0>,
\label{fulstateeven}
\end{eqnarray}
or any state which follows from one of the states of Eq.(\ref{fulstateeven}) by replacing any of  
the operators $\;(\gamma^a \pm \sqrt{-\eta^{aa} \eta^{bb}}\; \gamma^b)\;$ by the corresponding
$\;(1 \pm \sqrt{-\eta^{aa}\eta^{bb}} \;\gamma^a \gamma^b)\;$ are the eigenstates of all the operators
of the Cartan subalgebra of Eq.(\ref{choicecartan}) for an even dimension, 
with the eigenvalue of the chosen $S^{ab}$
determined by {\em Theorem 1}, while the eigenvalue of the operator of handedness (Eq.(\ref{hand})) 
can easily be calculated by taking into account Eqs.(\ref{hand},\ref{chequeeigensab}) and the fact that
$ \gamma^a \gamma^b = -2i S^{ab}$.

For an odd dimension ($d=2n+1$) the states
\begin{eqnarray}
(1 \pm  \Gamma) \gamma^0 (\gamma^1 \pm \sqrt{-\eta^{11} \eta^{22}}\; \gamma^2)\;(\gamma^3 
\pm \sqrt{-\eta^{33} \eta^{55}}\; \gamma^5)
\cdots (\gamma^{d-1} \pm \sqrt{-\eta^{d-1 d-1} \eta^{d d}}\; \gamma^{d})|\psi_0>,
\label{fulstateodd}
\end{eqnarray}
or any state which follows from  one of the states of Eq.(\ref{fulstateodd}) by replacing any of  
the operators $(\gamma^a \pm \sqrt{-\eta^{aa} \eta^{bb}}\; \gamma^b)$ by the corresponding
$(1 \pm \sqrt{-\eta^{aa}\eta^{bb}} \;\gamma^a \gamma^b)$, are the eigenstates of all the operators
of the Cartan subalgebra of Eq.(\ref{choicecartan}) of odd dimension, with the eigenvalues 
of the chosen $S^{ab}$ determined in {\em Theorem 1} and with the operator $\Gamma$ defined in Eq.(\ref{hand}). 

We assume that the ``vacuum state'' $|\psi_0>$ is chosen in such a way that the  above
operators, which are polynomials of $\gamma^a$'s, do not transform the ``vacuum state'' into zero.

The proof for $d$ even or $d$ odd follows from {\em Theorem 1} and the fact that $\Gamma$ commutes with all $S^{ab}$.

It is simple to count the number of basic states, which is $2^d$ (1 and $\gamma^a$ for each coordinate $a$).

\section{Irreducible representations of Lorentz group}
\label{irreducible}

To find  irreducible representations of the Lorentz group we only need to start with one 
of the proposed vectors. All the other members of the same irreducible representation follow by 
application of the generators of the 
Lorentz group $S^{ab}$, which do not belong to the Cartan subalgebra (Eq.(\ref{choicecartan}))
(or by applying on a chosen state a group element $O(\omega) = \exp{(-(i/2) \omega_{ab} S^{ab})}\;\;$).

We present a theorem, which helps to find all the irreducible representations of the Lorentz group.

{\em Theorem 3: } Let $S^{ab}$ and $S^{cd}$ be the two elements of the Cartan subalgebra. Then the
two vectors 
\begin{eqnarray}
(\gamma^a + \sqrt{-\eta^{aa} \eta^{bb}}\;\; \gamma^b)\; (\gamma^c + \sqrt{-\eta^{cc}\eta^{dd}} \;\;
\gamma^d)|\psi>,
\nonumber\\
(1 + \frac{1}{\eta^{aa}} \sqrt{-\eta^{aa} \eta^{bb}}\;\; \gamma^a \gamma^b) \;  (1 + \frac{1}{\eta^{cc}}
\sqrt{-\eta^{cc}\eta^{dd}} \;\;\gamma^c \gamma^d)|\psi>
\label{eigenirred}
\end{eqnarray}
belong to the same irreducible representation, where $|\psi>$ is any state, which the above operator
does not transform into zero.

{\em Proof:} To prove the theorem we apply the operator $S^{ac}=(i/2) \gamma^a \gamma^c$ 
(or the operator $S^{ad}$ or $S^{bc}$ or $S^{bd}$) to the first state
\begin{eqnarray}
\frac{i}{2} \gamma^a \gamma^c \;\;(\gamma^a + \sqrt{-\eta^{aa} \eta^{bb}}\;\; \gamma^b)\; 
(\gamma^c + \sqrt{-\eta^{cc}\eta^{dd}}\;\; \gamma^d)|\psi> =
\nonumber\\
 \- \frac{i}{2} \eta^{aa} \eta^{cc}(1 + \frac{1}{\eta^{aa}} \sqrt{-\eta^{aa} \eta^{bb}}\;\; 
 \gamma^a \gamma^b) \;  (1 + \frac{1}{\eta^{cc}}\sqrt{-\eta^{aa}\eta^{bb}}\;\; \gamma^c \gamma^d)|\psi>.
\label{eigenirredproof}
\end{eqnarray}

{\em Theorem 4:}
In an odd-dimensional space  the two states
\begin{eqnarray}
(1 \pm \Gamma)\gamma^0\;\;(\gamma^1 + \sqrt{-\eta^{11} \eta^{22}}\;\; \gamma^2) (\gamma^3 + 
\sqrt{-\eta^{33}\eta^{55}}\;\;\gamma^5)\cdots|\psi>,
\nonumber\\
(1 \pm \Gamma)(\gamma^1 +  \sqrt{-\eta^{11} \eta^{22}} \gamma^2)(\gamma^3 + \sqrt{-\eta^{33}\eta^{55}}\gamma^5)
\cdots|\psi>
\label{eigenirredodd}
\end{eqnarray}
are proportional to each other.

{\em Proof:} We first notice that $\Gamma = :\Gamma^{(d)} = \sqrt{\eta^{00}}\; \gamma^0\; \Gamma^{(d-1)}. $
The indices $(d)$ and $(d-1)$ were added to point out that $\Gamma^{(d-1)}$ includes all the $\gamma^a$'s, 
except the first one with factors which guarantee that either $\Gamma^{(d)}$ or $\Gamma^{(d-1)}$
fulfills the conditions of Eq.(\ref{prophand}). Then we see that when $\Gamma^{(d)}$ is applied to 
the state $\gamma^0(\gamma^1 + \sqrt{-\eta^{11} \eta^{22}}\; 
\gamma^2) (\gamma^3 + \sqrt{-\eta^{33}\eta^{55}}\; \gamma^5)\cdots|\psi_0>,$ gives $\sqrt{\eta^{00}}\eta^{00}$ 
$\varepsilon (\gamma^1 + \sqrt{-\eta^{11} \eta^{22}}\; 
\gamma^2) (\gamma^3 + \sqrt{-\eta^{33}\eta^{55}}\; \gamma^5)\cdots|\psi_0>,$ where 
 $\varepsilon (= \pm 1)$ is the eigenvalue of the operator $\Gamma^{(d-1)}$ on the chosen state.
 Then it follows $(1 \pm \Gamma)\gamma^0 (\gamma^1 + \sqrt{-\eta^{11} \eta^{22}}\; 
\gamma^2) (\gamma^3 + \sqrt{-\eta^{33}\eta^{55}}\; \gamma^5)\cdots|\psi_0> = \varepsilon \eta^{00} \sqrt{\eta^{00}}
 (1 \pm \Gamma)(\gamma^1 + \sqrt{-\eta^{11} 
\eta^{22}}\; \gamma^2) (\gamma^3 + \sqrt{-\eta^{33}\eta^{55}}\; \gamma^5)\cdots|\psi_0>$. This completes the proof.

As a consequence of {\em Theorem 4} the following statement follows

{\em Statement 1:} In odd-dimensional spaces $\gamma^a$'s do not transform one irreducible representation of
the Lorentz group into another as they do in even-dimensional spaces.

The proof is self-evident if we take account of {\em Theorem 4} and the fact that $\gamma^0 \gamma^a = -2i S^{0a}$.
(From {\em Theorem 3} and {\em Theorem 4} it follows that by applying to a chosen state the
operators $S^{01}, S^{03},\cdots, S^{0 d-2},S^{13},\cdots,S^{1 d-2}, \cdots, $ for an even $d$
and the operators $S^{01}, S^{05},\cdots, S^{0 d-1},S^{15},\cdots,S^{1 d-1}, \cdots, $ for an odd
$d$, we obtain all the members of a particular irreducible representation.)

{\em Statement 2a:} We find accordingly for $d$ even $2^{d/2-1}$ and for $d$ odd $2^{(d+1)/2 -1}$ members 
of an irreducible  representation, 
which is either left ($<\Gamma> = -1$) or right  ($<\Gamma> = +1$) handed. 

We call these representations the Weyl spinor representations. 
In section (\ref{graphic}) we present the
graphic way of looking for the representations.
 We can count the number of states by counting the number of $S^{ab}$ and products of $S^{ab}$
(not belonging to the Cartan subalgebra) which, when applied to a chosen state, transform this state into new 
states.

{\em Statement 2b:}
By applying any of $\gamma^a$'s to any of the states belonging to the above-obtained irreducible representation
of the chosen handedness for an even $d$, the corresponding Weyl irreducible representation of the opposite 
handedness follows. 

Since $\Gamma$ commutes with  $S^{ab}$ and in even $d$ anticommutes with $\gamma^a$, 
the change of handedness,  when $\gamma^a$ is  used to obtain a
new irreducible representation from the old one, is self-evident.

Two Weyl spinors of the opposite handedness together have $2^{d/2}$ states for  even $d$. 

We all the time think of $\gamma^a$'s as formal operators, not as matrices.  
In the construction of eigenstates according to {\em Theorem 2}  
(Eq.(\ref{fulstateeven})) one can of course think of the operator $\gamma^a$ as a matrix, working on
just a certain space of a (single) Dirac spinor - what one usually does. In this case 
the state, constructed by either the nilpotent operator $(\gamma^a \pm \sqrt{-\eta^{aa}
\eta^{bb}} \gamma^b )$ or by the corresponding ``projection'' operator
$ ( 1 \pm \sqrt{-\eta^{aa}
\eta^{bb}} \gamma^a \gamma^b)$, represent up to a sign
$\pm$ the same eigenstate of the Cartan operator $S^{ab}$. So if these matrices work on
just a certain space of Dirac spinors and there is only one component with 
a given eigenvalue combination of the $d/2$ or $(d-1)/2$  Cartan algebra matrices 
in respectively even and odd $d$
cases, we must get proportional eigenstates. Different 
constructions, of which we can make $2^{d}$ states, 
are thus in the case, when only one Dirac spinor is concerned, far from being
linearly independent. In fact there are in the even case $2^{d/2}$ and in the odd case $2^{(d+1)/2}$
different formal products of the type of {\em Theorem 2} which all lead to the same 
eigenstates. 

We might, however, decide to play the formal game that these 
different factors in our {\em Theorem 2} lead to linearly independent 
states. 
In fact such a case can be realized (section \ref{inner}), if we use as the the ``vacuum state'' 
$|\psi_0>$ not  a single Dirac spinor but rather  $2^{d/2}$ Dirac spinors 
for even and $2^{(d+1)/2}$ Dirac spinors for odd $d$, and put them
into a direct sum. In the generic case we would in this case achieve 
the suggested linear independence.
So, if we assume that there are no linear relations 
among  states constructed by means of our nilpotent and 
``projection'' operators  - more than the ones that follow from the 
Clifford algebra, which in turn means the projection and nilpotency 
relations and the commutation rules - we would get the 
$2^{d}$ now assumed independent states.
Since we know that the irreducible representation of 
a $\gamma^a$ or Clifford algebra is only $2^{d/2}$ dimensional,
the $2^{d}$ linearly independent - by formal game -
states must divide themselves into $2^{d}/2^{d/2}$ = $2^{d/2}$ for $d$ even and $2^d/ 2^{(d-1)/2} =
2^{(d+1)/2}$ for $d$ odd,
different ``families''\cite{norma94,norma99,holgernormadk}.

For $d$ even we find  $2^{d/2}$ ``families''  by looking for states not
yet included in the starting two Weyl representations  and repeating the procedure for finding from  starting 
states all the states belonging to an irreducible representation.

 For $d$ odd half of the states (of $2^{(d-1)/2}$ members of an irreducible representation of the Lorentz 
group ) are obtained by applying $(1 + \Gamma)$ to the chosen eigenstate of all the Cartan generators, 
and the other half by applying  $(1 - \Gamma)$ to the same chosen eigenstate. Again 
we find all $2^{(d+1)/2}$``families'' by looking for a state which is not yet a member of the Weyl spinor irreducible
representations already found and by repeating the procedure for finding from a starting 
states all the states belonging to an irreducible representation.

\section{Inner product and ortonormalization of basic states}
\label{inner}

We presented in section \ref{irreducible} a simple technique  to find states which belong to one 
irreducible representation of the Lorentz group, 
and we also demonstrated, when playing the game that the nilpotent and the ``projection'' operator form  linear 
independent states, that there are in even-dimensional spaces  $2^{d/2}$ ``families'' of the Weyl 
two-spinors  (that is one Dirac spinor) and in odd-dimensional spaces 
$2^{(d+1)/2}$ ``families'' of the Weyl one-spinors. We shall prove in this section that the  proposed states, 
generated by applying the appropriately chosen polynomials of the operators $\gamma^a$'s to the 
``vacuum state''  are orthonormal if the vacuum state is properly chosen.

{\em Theorem 5:} Let  the ``vacuum state''  $ |\psi_0>$   be the superposition of all spinor states 
belonging to one Weyl ``family'' with coefficients $\alpha_i, 
i=1,\cdots,  2^{d/2}$ having the property $\alpha_i^* \alpha_i = 1/2^{d/2}$, for $d$ even
and $\alpha_i^* \alpha_i = 1/ 2^{(d-1)/2},$ $i=1,\cdots,  2^{(d-1)/2}$, for $d$ odd. 
Then this ``vacuum state'' being normalized 
\begin{eqnarray}
<\psi_0|\psi_0> = 1
\label{psi0}
\end{eqnarray}
has as a consequence that all  the states presented in section \ref{irreducible} as belonging 
to one ``family'' are orthogonal and, if the states are multiplied by $1/\sqrt{2^r}$, where $r$ ($r = d/2$ 
for $d$ even and $r= (d-1)/2$ for $d$ odd) is the 
number of the operators in the Cartan subalgebra, also orthonormal.

{\em Proof:} To prove this theorem we first start with an even $d=2n$ and show that if the ``vacuum state''
$|\psi_0>$  fulfills the above-required condition, any state, which is the eigenstate of the generators of 
the Cartan subalgebra of the Lorentz group, can be normalized.
We  take the state (see Eq.(\ref{fulstateeven}))
\begin{eqnarray}
|{}^1\psi_1> = \frac{1}{\sqrt{2}}\;(\gamma^0 + \sqrt{-\eta^{00} \eta^{dd}}\;\;
\gamma^d)\;\frac{1}{\sqrt{2}}\;(\gamma^1 + \sqrt{-\eta^{11} \eta^{22}}\; \;\gamma^2)
\cdots \frac{1}{\sqrt{2}} \;(\gamma^a + \sqrt{-\eta^{aa}\eta^{bb}}\; \; \gamma^b)\;
\nonumber\\
\cdot\frac{1}{\sqrt{2}}\; (\gamma^c + \sqrt{-\eta^{cc}\eta^{ee}}\; \gamma^e)\;\cdots
\frac{1}{\sqrt{2}}\;(\gamma^{d-2} + \sqrt{-\eta^{d-2 d-2} \eta^{d-1 d-1}}\;\; \gamma^{d-1})|\psi_0>,
\label{orthogstateevennor}
\end{eqnarray}
and find the inner product $\;<{ }^1\psi_1|{ }^1\psi_1>$. Let us look first at the  inner most 
part of the operators:
$\;(1/2)\;((\gamma^0)^{\dagger} + (\sqrt{-\eta^{00} \eta^{dd}}\;\gamma^d)^{\dagger})\;
(\gamma^0 + \sqrt{-\eta^{00} \eta^{dd}}
\;\gamma^d)$.
Multiplying the two binomials we find $\;(1 +  \sqrt{-\eta^{00}\eta^{dd}}\;\eta^{00}\;\gamma^0 \gamma^d)$. 
Each of the
two corresponding binomials, defining the eigenstate of the particular member of the Cartan subalgebra $S^{ab}$,
will contribute a similar part: $(1 +  \sqrt{-\eta^{aa}\eta^{bb}}\;\eta^{aa}\;\gamma^a \gamma^b)$. The product
of all unit operators will give $<\psi_0|\psi_0>$. The product of each of $\gamma^a \gamma^b$ with the unit
operators will give $<\psi_0|\gamma^a \gamma^b|\psi_0>= <\psi_0|-2iS^{ab}|\psi_0>$, for all members 
$S^{ab}$ of the chosen Cartan subalgebra. This gives zero due to the above-declared choice of 
the ``vacuum state'' $|\psi_0>$. We shall also obtain the product of four two members of the Cartan subalgebra
$<\psi_0|(-2i)S^{ab}(-2i)S^{ce}|\psi_0>$ and so on, but all the expectation values of this kind, due to the
construction of $|\psi_0>,$ are equal to zero. We can conclude that each  state, which is the eigenstate of 
all the members of the chosen Cartan subalgebra, is normalized to unity, if each binomial $(1/\sqrt{2})
(\gamma^a + \sqrt{-\eta^{aa} \eta^{bb}}\; \gamma^b)$ appears with factor of normalization $1/\sqrt{2}$.

We have now to look for the orthogonality properties of two states belonging to the same 
irreducible representation, where one state is obtained from the other by the application of a generator
$S^{ac}$ of the Lorentz group 
which is not the member of the Cartan subalgebra. Such two states are the state of Eq.(\ref{orthogstateevennor}) 
and the state
\begin{eqnarray}
|{}^1\psi_2> = \eta^{aa}\eta^{cc}\;\frac{1}{\sqrt{2}}\;(\gamma^0 + \sqrt{-\eta^{00} \eta^{dd}}\; \gamma^d)\;\;
\frac{1}{\sqrt{2}}(\gamma^1 + \sqrt{-\eta^{11} \eta^{22}}\;\; \gamma^2)
\cdots \nonumber\\
\cdots
\frac{1}{\sqrt{2}}(1 + \frac{1}{\eta^{aa}}\sqrt{-\eta^{aa}\eta^{bb}} \;\;\gamma^a \gamma^b)
\frac{1}{\sqrt{2}}(1 + \frac{1}{\eta^{cc}}
\sqrt{-\eta^{cc}\eta^{ee}} \; \;\gamma^c \gamma^e)
\cdots \nonumber\\
\cdots
\frac{1}{\sqrt{2}}(\gamma^{d-2} + \sqrt{-\eta^{d-2 d-2} \eta^{d-1 d-1}}\; \gamma^{d-1})|\psi_0>.
\label{orthogostateevenorto}
\end{eqnarray}
The bra $<{ }^1\psi_1|$ differs from the ket $|{ }^1\psi_2>$ in the product of the two binomials, the 
first gives with the corresponding partner from the right the product 
$(\gamma^a + \sqrt{-\eta^{aa}\eta^{bb}}\;\gamma^b)^{\dagger}(1 + (1/\eta^{aa}) \sqrt{-\eta^{aa}\eta^{bb}}\;
\;\gamma^a \gamma^b) = 0.$ Similarly, the second factor, when multiplied by the corresponding partner 
from the right, gives zero, and we conclude that the two states appearing in the bra and the ket are orthogonal.

The application of any of operators $\gamma^a$ (which is not a member of the Lorentz algebra)
takes us from one Weyl spinor representation of one 
handedness to another of the opposite handedness. We only have to apply any $\gamma^a$
on any state of the starting  handedness (and then the generators of the Lorentz algebra generate all the
members of the irreducible representation of the opposite handedness). We have therefore to check the 
orthogonality relation between such two states. We shall start with the state of 
 Eq.(\ref{orthogstateevennor}) and the state 
\begin{eqnarray}
|{}^2\psi_1> = \eta^{aa}\;\frac{1}{\sqrt{2}}(\gamma^0 + \sqrt{-\eta^{00} \eta^{dd}}\;\; \gamma^d)\;
\frac{1}{\sqrt{2}}(\gamma^1 + \sqrt{-\eta^{11} \eta^{22}}\; \gamma^2)
\cdots 
\nonumber\\
\frac{1}{\sqrt{2}}(1 + \frac{1}{\eta^{aa}}\; \sqrt{-\eta^{aa}\eta^{bb}} \; \;\gamma^a \gamma^b)
\cdots
\frac{1}{\sqrt{2}}(\gamma^{d-2} + \sqrt{-\eta^{d-2 d-2} \eta^{d-1 d-1}}\; \gamma^{d-1})|\psi_0>.
\label{orthogostateevenortogamma}
\end{eqnarray}

The factor which determines the orthogonality of the two states is $(\gamma^a + \sqrt{-\eta^{aa}\eta^{bb}}
\; \gamma^b)^{\dagger}
(1 + (1/\eta^{aa})\sqrt{-\eta^{aa}\eta^{bb}} \;\gamma^a \gamma^b)$. 
Multiplying the two binomials and summing the corresponding terms, one finds that it is zero.

To further prove the orthogonality relations for odd-dimensional spaces ($d=2n+1$), we follow the procedure, 
presented for even-dimensional spaces with the requirement that  the binomial
$(1 + \Gamma)$ appears with the same normalizing factor $1/\sqrt{2}$ as all the other binomials.

We conclude the proof  with the statement that each state $|{ }^h \psi_i>$ belonging to an irreducible 
representation of handedness  $h$, is, due to 
appropriately chosen ``vacuum state'' and the normalization factor (which is $1/\sqrt{2}$ 
for each binomial) orthonormal to any other state $|{ }^{h'} \psi_j>$
\begin{eqnarray}
<{ }^h \psi_i|{ }^{h'} \psi_j> = \delta^{h h'}\delta^{ij}.
\label{orthogonalfamily}
\end{eqnarray}

We would like to point out that if we treat Weyl spinors of one handedness only, then the ``vacuum state''
should  normalize to $1/2$ rather than to $1$, since then the product of all members of the Cartan 
subalgebra, multiplied by the appropriate factor and applied to such a ``vacuum state'' will give one instead of zero.

\section{``Families'' of Lorentz group}
\label{families}

Taking into account the orthogonality of all the $2^d$ polynomials of the $\gamma^a$ operators and assuming that 
these polynomials act on a ``vacuum state'' which assures the orthogonality of all the obtained states we obtained 
by counting the number of all states, which is for $d$-dimensional space $2^d$, and comparing  for $d$ even 
this number with twice the number of states 
in one Weyl spinor irreducible representation  (since we count states of both handedness), 
which is $2 \times 
2^{d/2-1}$, we found $2^{d/2}$ copies of the two-Weyl spinors (that is of the Dirac spinord). For $d$ odd we 
found accordingly $2^{(d+1)/2}$
copies of the Weyl spinor. We  call these $2^{d/2}$ copies for $d$ even and $2^{(d+1)/2}$
copies for $d$ odd ``families'' of spinors.
Each ``family''  differs from all the others in the choice of the starting state, 
on which the irreducible representation is built, and accordingly on all the states. 

To achieve the required orthogonality, we make   a choice of 
phases for states belonging to different ``families'' in such a way that, when choosing the ``vacuum state'' $|\psi_0>$ 
 to be the sum of not only 
all the states in one ``family'' but of all the states of all the ``families'', each state appearing with the same 
coefficient, not only the expectation values of the operators $S^{ab}$, belonging
to the Cartan subalgebra, but also the expectation values of $\gamma^a$ are for  such $|\psi_0>$ equal to zero, 
that is $<\psi_0|\gamma^a|\psi_0> =0$, for $d$ even and odd. 
Given such a choice of the ``vacuum state'' all states belonging to any ``family'' are orthonormal
\begin{eqnarray}
<{ }^a \psi_i|{ }^{b} \psi_j> = \delta^{ab}\delta^{ij}.
\label{othonormalfamilies}
\end{eqnarray}
We use index $a$ or $b$ to numerate a ``family'' and index $i$ or $j$ to numerate states within a ``family''.

\section{Graphic presentations of the Lorentz group}
\label{graphic}

We shall present in this section a simple and transparent graphic technique for finding irreducible 
representations of the Lorentz group for spinors for any dimension - even or odd. We start by introducing 
the notation (already seen in {\em Theorem 1})
\begin{eqnarray}
\stackrel{ab}{(k)}:&=&\frac{1}{\sqrt{2}}(\gamma^a + \frac{\eta^{bb}}{i}k\gamma^b) = 
\frac{1}{\sqrt{2}}(\gamma^a + \frac{\eta^{aa}}{ik}\gamma^b)\nonumber\\
\stackrel{ab}{[k]}:&=&\frac{1}{\sqrt{2}}(1+k\eta^{aa}\eta^{bb}i\gamma^a\gamma^b)\nonumber\\
\stackrel{+}{\circ}:&=& \frac{1}{\sqrt{2}} (1+\Gamma)\nonumber\\
\stackrel{-}{\bullet}:&=& \frac{1}{\sqrt{2}}(1-\Gamma)
\label{signature}
\end{eqnarray}
under the assumption that the eigenvalue $k$ of $2S^{ab}$ ( supposedly one of the Cartan-algebra generators)
being of course restricted by $k^2 = \eta^{aa}\eta^{bb}$, since it is real for $a$ and $b$ corresponding
both to time or both to space index, while it is purely imaginary for opposite signatures. We see that 
{\em Theorem 1} above tells us
\begin{eqnarray}
S^{ab}\stackrel{ab}{(k)}=\frac{1}{2}k\stackrel{ab}{(k)}\nonumber\\
S^{ab}\stackrel{ab}{[k]}=\frac{1}{2}k\stackrel{ab}{[k]}.
\label{grapheigen}
\end{eqnarray}
We have of course for $d$ different from $a$ and $b$
\begin{eqnarray}
\gamma^d \stackrel{ab}{(k)}=-\stackrel{ab}{(k)} \gamma^d \nonumber\\
\gamma^d \stackrel{ab}{[k]}=\stackrel{ab}{[k]} \gamma^d.
\label{graphcom}
\end{eqnarray}
We can easily find 
\begin{eqnarray}
\gamma^a \stackrel{ab}{(k)}&=&\eta^{aa}\stackrel{ab}{[-k]},\nonumber\\
\gamma^b \stackrel{ab}{(k)}&=& -ik \stackrel{ab}{[-k]}, \nonumber\\
\gamma^a \stackrel{ab}{[k]}&=& \stackrel{ab}{(-k)},\nonumber\\
\gamma^b \stackrel{ab}{[k]}&=& -ik \eta^{aa} \stackrel{ab}{(-k)}.
\label{graphgammaaction}
\end{eqnarray}

From Eqs.(\ref{graphcom},\ref{graphgammaaction}) it follows that
\begin{eqnarray}
S^{ac}\stackrel{ab}{(k)}\stackrel{cd}{(k)}& = &-\frac{i}{2} \eta^{aa} \eta^{cc} 
\stackrel{ab}{[-k]}\stackrel{ab}{[-k]} \nonumber\\
S^{ac}\stackrel{ab}{[k]}\stackrel{cd}{[k]}& = &\frac{i}{2}  
\stackrel{ab}{(-k)}\stackrel{ab}{(-k)} \nonumber\\
S^{ac}\stackrel{ab}{(k)}\stackrel{cd}{[k]}& = &-\frac{i}{2} \eta^{aa}  
\stackrel{ab}{[-k]}\stackrel{ab}{(-k)}.
\label{sac}
\end{eqnarray}

We also find
\begin{eqnarray}
\stackrel{ab}{(k)}^{\dagger}&=&\eta^{aa}\stackrel{ab}{(-k)},\\
\stackrel{ab}{[k]}^{\dagger}&=& \stackrel{ab}{[k]}.
\label{graphher}
\end{eqnarray}
With respect to normalization, one may make slightly different proposals:
With the $1/2$-normalization written above, the states we produce by acting 
successively with the proposed operators would be normalized, provided that the ``vacuum state''
$|\psi_0>$ had equal probability for all its Cartan-algebra common eigenvalues
( as proposed above), but if we instead used a factor $1/2$ instead of  
$1/\sqrt{2}$, the $\stackrel{ab}{[k]}$ operators would 
be normalized as projection operators as they are. This would then require that we used
a $|\psi_0>$ state which was normalized so that only the eigenstate which we use 
for the considered choice of using nilpotents and projections is by itself normalized.
That is to say, this choice of normalization of the operators $\stackrel{ab}{[k]}$, and
necessarily also $\stackrel{ab}{(k)}$, requires that we would only  
have a normalized $|\psi_0>$ if it had all its probability concentrated on the single 
Cartan-algebra eigenstate components  used with the chosen combination of 
projections and nilpotents.

Let us conclude this section by presenting graphically a Weyl spinor irreducible representation
for $d$-dimensional space, with $d$ even and for one of $2^d$ possible ``families''. All the pairs
$S^{ab}, S^{cd},\cdots $, are members of the Cartan subalgebra of the Lorentz group.
\begin{eqnarray}
\stackrel{ab}{(k_{ab})} \stackrel{cd}{(k_{cd})} \stackrel{ef}{(k_{ef})}\cdots \stackrel{gh}{(k_{gh})}
\cdots |\phi_0> \nonumber\\
\eta^{aa}\eta^{cc}\stackrel{ab}{[-k_{ab}]} \stackrel{cd}{[-k_{cd}]} \stackrel{ef}{(k_{ef})}\cdots \stackrel{gh}{(k_{gh})}
\cdots |\phi_0> \nonumber\\
\eta^{aa}\eta^{ee}\stackrel{ab}{[-k_{ab}]} \stackrel{cd}{(k_{cd})} \stackrel{ef}{[-k_{ef}]}\cdots \stackrel{gh}{(k_{gh})}
\cdots |\phi_0> \nonumber\\
\vdots \nonumber\\
\eta^{aa}\eta^{gg}\stackrel{ab}{[-k_{ab}]} \stackrel{cd}{(k_{cd})} \stackrel{ef}{(k_{ef})}\cdots \stackrel{gh}{[-k_{gh}]}
\cdots |\phi_0> \nonumber\\
\eta^{cc}\eta^{ee}\stackrel{ab}{(k_{ab})} \stackrel{cd}{[-k_{cd}]} \stackrel{ef}{[-k_{ef}]}\cdots \stackrel{gh}{(k_{gh})}
\cdots |\phi_0> \nonumber\\
\vdots \nonumber\\
\label{graphicd}
\end{eqnarray}
What we learn from the above graphic representation is that 
one obtains all states of an irreducible Weyl representation by transforming all possible pairs
of $\stackrel{ab}{(k_{ab})} \stackrel{mn}{(k_{mn})}$ to $\stackrel{ab}{[-k_{ab}]} \stackrel{mn}{[-k_{mn}]}$.
The procedure gives $2^{(d/2-1)}$ states. We shall use the  presented graphical search to find the Weyl spinors 
for $d=3$ and $d=4$ in
the next section.

\section{Demonstration of irreducible representations of Weyl  spinors.  }
\label{demonstration}

In this section we demonstrate what we have learned, on two cases: We look for the
irreducible representations of spinors with respect to the Lorentz
group for a three-dimensional  and a four-dimensional case. In both cases we shall assume the Minkowski
metric: $\eta^{00} = -\eta^{ii} $, with $i =1,2$ for $d=3$ and $i=1,2,3$ for $d=4$.
We point out again, that the ``vacuum state'' is chosen in such a way that all $2^d$ linearly independent polynomials
of the $\gamma^a$ operators form, when applied to the ``vacuum state'', $2^d$ orthogonal basic vectors, which 
are also normalized.

\subsection{``Families'' of Weyl spinors for $d=3$}
\label{three}

There is only one $((d-1)/2=1)$ operator of the Cartan subalgebra. According to Eq.(\ref{choicecartan})
we choose $S^{12}$ as the member of the Cartan subalgebra of the Lorentz algebra which  the
operators $S^{01}, S^{02}, S^{12}$ close. Following Eq.(\ref{hand}) we find $\Gamma = 
i \gamma^0 \gamma^1 \gamma^2. $ There are $2^3$, that is eight, basic states, which we arange 
to be eigenstates of $S^{12}$ 
\begin{eqnarray}
\frac{1}{2}(1 \pm \Gamma) (\gamma^1 \pm i \gamma^2), \quad \frac{1}{2}\;\gamma^0\;(1 \pm \Gamma) 
(1 \pm i \gamma^1 \gamma^2)=\frac{1}{2}\;\varepsilon (1\pm \Gamma)(1 \pm i\gamma^1 \gamma^2),
\label{basisthree}
\end{eqnarray}
with $\varepsilon =1,$ if in both factors $(1\pm \Gamma)$ and $(1\pm \gamma^1 \gamma^2)$ the same sign (either 
$++$ or $--$) is taken and $\varepsilon = -1,$ otherwise (for the cases $+-$ or $-+$). 
We find the eigenvalues of the Cartan operator $S^{12}$ according to {\em Theorem 1} for the eight states of
Eq.(\ref{basisthree})
to be $\pm 1/2$.\\
We arrange these eight orthonormal states into four ``families'' as presented in Table I.

\begin{center}
\begin{tabular}{|r|r||c||r||c|}
\hline
a&i&$|^a\psi_i>$&$ S^{12}$& graphic presentation \\
\hline\hline
1&1&$\frac{1}{2}(1 + \Gamma) (\gamma^1 + i\gamma^2)|\psi_0>$& $\frac{1}{2}$ &$\stackrel{+}{\circ} \; 
\stackrel{12}{(+)}$ or $\circ (+)$  \\
\hline 
1&2&$\frac{1}{2}(1 + \Gamma)(1 -i\gamma^1 \gamma^2)|\psi_0>$ & $-\frac{1}{2}$&$\stackrel{+}{\circ}
\; \stackrel{12}{[-]}$ or $ \circ [-]$\\
\hline \hline
2&1&$\frac{1}{2} (1 - \Gamma)(\gamma^1 + i \gamma^2)|\psi_0>$& $\frac{1}{2}$&$\stackrel{-}{\bullet} 
\; \stackrel{12}{(+)}$ or $\bullet (+)$\\
\hline 
2&2&$\frac{1}{2}(1 - \Gamma) (1 -i \gamma^1 \gamma^2)|\psi_0>$& $-\frac{1}{2}$&$\stackrel{-}{\bullet} 
\; \stackrel{12}{[-]}$ or $\bullet [-]$\\
\hline\hline 
3&1&$-\frac{1}{2}(1 + \Gamma)(1 + i \gamma^1 \gamma^2)|\psi_0> $& $\frac{1}{2}$&$\stackrel{+}{\circ}
\; \stackrel{12}{[+]}$ or $\circ [+]$ \\
\hline 
3&2&$-\frac{1}{2}(1 + \Gamma)(\gamma^1 - i \gamma^2)|\psi_0> $& $-\frac{1}{2}$&$-\stackrel{+}{\circ}
\; \stackrel{12}{(-)}$ or $-\circ (-)$\\
\hline \hline
4&1&$-\frac{1}{2}(1 - \Gamma)( 1 + i \gamma^1 \gamma^2)|\psi_0> $& $\frac{1}{2}$&$\stackrel{-}{\bullet} 
\; \stackrel{12}{[+]}$  or $\bullet [+]$\\
\hline 
4&2&$-\frac{1}{2}(1 - \Gamma)(\gamma^1 - i \gamma^2)|\psi_0> $& $-\frac{1}{2}$&$\stackrel{-}{\bullet} 
\; \stackrel{12}{(-)}$ or $\bullet (-)$\\
\hline\hline 
\end{tabular}
\end{center}

Table I.- Irreducible representations of the Lorentz group  $SO(1,2)$, arranged into four ``families''. All 
vectors are eigenvectors of $S^{12}$ and are orthonormalized in the sense discussed in section (\ref{inner}).
The graphical presentation follows the procedure 
described in section (\ref{graphic}). We add also a simplified version of the graphical presentation of states.

Any of the four ``families'' can be used as the basis when solving the massive or massless Dirac equation,
as it will be demonstrated in next section (\ref{solutions}).

\subsection{``Families'' of Weyl spinors for $d=4$}
\label{three}

There are two  $(d/2=2)$ operators of the Cartan subalgebra of the Lorentz algebra closed by the operators 
$S^{01}, S^{02}, S^{03}, S^{12}, S^{13}, S^{23}$. According to 
Eq.(\ref{choicecartan})
we choose $ S^{03}$ and $S^{12}$ as the members of the Cartan subalgebra. Following Eq.(\ref{hand}) 
we find $\Gamma = 
i \gamma^0 \gamma^1 \gamma^2 \gamma^3. $ There are $2^4$ , that is sixteen basic states,
all of them  being  eigenstates of $S^{12}$ and $S^{03}$
\begin{eqnarray}
\frac{1}{2}(\gamma^0 \pm \gamma^3) (\gamma^1 \pm i \gamma^2), \quad \frac{1}{2}(\gamma^0 \pm \gamma^3) 
(1 \pm i \gamma^1 \gamma^2),
\nonumber\\
\frac{1}{2}(1 \pm \gamma^0 \gamma^3) (\gamma^1 \pm i \gamma^2), \quad\frac{1}{2}(1 \pm \gamma^0
\gamma^3) (1 \pm i \gamma^1 \gamma^2).
\label{basisfour}
\end{eqnarray}
According to {\em Theorem 1}, the eigenvalues of the Cartan operator $S^{12}$ for four times four   basic states of
Eq.(\ref{basisfour})
are $\pm 1/2$ and the eigenvalues of the Cartan operator $S^{03}$ for the four times four basic states
are  $\mp i/2$. All sixteen basic states are orthonormal.

We arrange these sixteen states into four ``families'' as presented in section (\ref{inner}). 
Each ``family'' includes two Weyl spinors, one left- and one right-handed. We come from one Weyl spinor, say left,
to another, say right, by applying $\gamma^a$'s to the left one.
These four ``families'' are  presented in Table II.

\begin{center}
\begin{tabular}{|r|r||c||r|r|r||c|}
\hline
a&i&$|^a\psi_i>$&$ S^{12}$&$ S^{03}$&$
\Gamma$&graphic presentation\\
\hline\hline
1&1&$\frac{1}{2}(\gamma^0 +
\gamma^3)(\gamma^1 + i \gamma^2) |\psi_0>$& $\frac{1}{2}$& $-\frac{i}{2}$& -1& $\quad \stackrel{03}
{(-i)} \stackrel{12}{(+)} $ or $(-i)(+)$\\
\hline 
1&2&$\frac{1}{2}(1+\gamma^0 
\gamma^3)(1 -i\gamma^1 \gamma^2) |\psi_0>$& $-\frac{1}{2}$& $\frac{i}{2}$& -1&$\quad\stackrel{03}
{[+i]} \stackrel{12}{[-]} $ or $[+i] [-]$\\
\hline 
2&1&$\frac{1}{2}(1 +\gamma^0 
\gamma^3)(\gamma^1 + i \gamma^2) |\psi_0>$& $\frac{1}{2}$& $\frac{i}{2}$& 1&$\quad\stackrel{03}
{[+i]} \stackrel{12}{(+)} $ or $[+i] (+)$\\
\hline 
2&2&$-\frac{1}{2}(\gamma^0+ 
\gamma^3)(1 -i\gamma^1 \gamma^2) |\psi_0> $& $-\frac{1}{2}$& $-\frac{i}{2}$& 1&$-\stackrel{03}
{(-i)} \stackrel{12}{[-]} $ or $-(-i) [-]$\\
\hline\hline 
3&1&$-\frac{1}{2}(1 -\gamma^0 
\gamma^3)(\gamma^1 + i \gamma^2) |\psi_0>$& $\frac{1}{2}$& $-\frac{i}{2}$& -1& $-\stackrel{03}
{[-i]} \stackrel{12}{(+)} $ or $-[-i] (+)$\\
\hline 
3&2&$\frac{1}{2}(\gamma^0 -
\gamma^3)(1 - i\gamma^1 \gamma^2) |\psi_0>$& $-\frac{1}{2}$& $\frac{i}{2}$& -1& $\quad \stackrel{03}
{(+i)} \stackrel{12}{[-]} $ or $(+i) [-]$\\
\hline
4&1&$\frac{1}{2}(\gamma^0-
\gamma^3)(\gamma^1 + i \gamma^2)|\psi_0>$& $\frac{1}{2}$& $\frac{i}{2}$& 1&$\quad \stackrel{03}
{(+i)} \stackrel{12}{(+)} $ or $(+i) (+)$\\
\hline 
4&2&$\frac{1}{2}(1-\gamma^0 
\gamma^3)(1 -i\gamma^1 \gamma^2) |\psi_0>$& $-\frac{1}{2}$& $-\frac{i}{2}$& 1&$\quad \stackrel{03}
{[-i]} \stackrel{12}{[-]} $ or $[-i] [-]$\\
\hline\hline 
5&1&$\frac{1}{2}(1 - \gamma^0
\gamma^3)(1 +i\gamma^1  \gamma^2) |\psi_0>$& $\frac{1}{2}$& $-\frac{i}{2}$& -1&$\quad \stackrel{03}
{[-i]} \stackrel{12}{[+]} $ or $[-i] [+]$\\
\hline
5&2&$\frac{1}{2} (\gamma^0 - 
\gamma^3)(\gamma^1 -i\gamma^2)|\psi_0>$& $-\frac{1}{2}$& $\frac{i}{2}$& -1&$\quad \stackrel{03}
{(+i)} \stackrel{12}{(-)} $ or $(+i) (-)$\\
\hline 
6&1&$\frac{1}{2}(\gamma^0 -
\gamma^3)(1 +i\gamma^1  \gamma^2) |\psi_0>$& $\frac{1}{2}$& $\frac{i}{2}$& 1&$\quad \stackrel{03}
{(+i)} \stackrel{12}{[+]} $ or $(+i) [+]$\\
\hline 
6&2&$-\frac{1}{2}(1-\gamma^0 
\gamma^3)(\gamma^1 -i \gamma^2) |\psi_0>$& $-\frac{1}{2}$& $-\frac{i}{2}$& 1& $-\stackrel{03}
{[-i]} \stackrel{12}{(-)} $ or $-[-i](-)$\\
\hline\hline 
7&1&$\frac{1}{2}(\gamma^0+ 
\gamma^3)(1 +i\gamma^1 \gamma^2) |\psi_0>$& $\frac{1}{2}$& $-\frac{i}{2}$& -1&$\quad \stackrel{03}
{(-i)} \stackrel{12}{[+]} $ or $(-i) [+]$\\
\hline 
7&2&$-\frac{1}{2}(1 +\gamma^0 
\gamma^3)(\gamma^1 - i\gamma^2) |\psi_0>$& $-\frac{1}{2}$& $\frac{i}{2}$& -1&$-\stackrel{03}
{[+i]} \stackrel{12}{(-)} $ or $-[+i] (-)$\\
\hline 
8&1&$-\frac{1}{2}(1 +\gamma^0 
\gamma^3)(1 +i\gamma^1 \gamma^2) |\psi_0>$& $\frac{1}{2}$& $\frac{i}{2}$& 1&$-\stackrel{03}
{[+i]} \stackrel{12}{[+]} $ or $-[+i] [+]$\\
\hline 
8&2&$-\frac{1}{2}(\gamma^0 + 
\gamma^3)(\gamma^1  -i\gamma^2) |\psi_0>$& $-\frac{1}{2}$& $-\frac{i}{2}$& 1&$-\stackrel{03}
{(-i)} \stackrel{12}{(-)} $ or $-(-i) (-)$\\
\hline\hline 
\end{tabular}
\end{center}
Table II.-Four ``families'' of the two Weyl spinors of the Lorentz group $SO(1,3)$.
Basic vectors are eigenfunctions of the two  operators of the Cartan subalgebra
$S^{12}$ and $S^{03}$. The eigenvalues of the operator of handedness $\Gamma$ are also 
presented. All the basic states are orthonormalized as discussed in section(\ref{inner}).
Two types of graphic presentation are added,
a simplified version in addition to the ordinary one.

Any of the four ``families'' can be used to present the solution of the Dirac  equation
for a massive spinor, while the massless spinors are either left- or right-handed, so that
only half of the space of the massive case is needed to find the solution.
We shall present  solutions of the Dirac equation for a massless or a massive case in the next section.

\section{Solutions of Weyl and Dirac equation for $d=3$ and $d=4$ }
\label{solutions}

We shall use the basic states presented in the previous section to find  solutions of the Dirac equation
for massless and massive cases in three-dimensional and four-dimensional spaces with one time ($\eta^{00} = 1$)
and $(d-1)$ space ($\eta^{ii}, i=1,..,d-1$ ) dimensions. We shall work, as usually one does, with only one 
family, making a choice of the first one 
in each of the two tables. Playing the formal game that nilpotent and ``projection'' 
operators give, if applied to the ``vacuum state'', linearly independent states, one finds four for $d$=$3$- 
and four for $d$ = $4$-dimensional case. The choice of any of the families in each case would be  equally good.

\subsection{Solution of the Dirac equation for  spinors in three-dimensional space}

To find the solution of the Dirac equation
\begin{eqnarray}
(\gamma^a p_a =m)|\psi>
\label{dirac3}
\end{eqnarray}
we shall use the basic states presented in Table I as the first ``family'' of two basic vectors.
Looking for the solution as a plane wave with wave vector $p^a = (p^0, p^1,p^2)$
we find that the state
\begin{eqnarray}
|\psi> = {\cal N}\frac{1}{2}(1 + \Gamma)\{\;(\gamma^1 + i \gamma^2) - 
\frac{p^1 + i p^2}{|p^0| + m}
 \;(1-i \gamma^1  \gamma^2)\}|\psi_0> e^{-ip^a x_a},
\label{dirac3sol}
\end{eqnarray}
or  graphically
\begin{eqnarray}
|\psi> = {\cal N}\;\{\;\stackrel{+}{\circ} \stackrel{12}{(+)} - \frac{p^1 + i p^2}{|p^0| + m}
\stackrel{+}{\circ} \stackrel{12}{[-]}\}\;|\psi_0> e^{-ip^a x_a},
\label{dirac3solgraph}
\end{eqnarray}
solves the Dirac equation for a massive case with  ${\cal N} = \sqrt{(|p^0| +m)/(2 p^0)}$ and 
$(p^{0})^2 = (p^{1})^2 + (p^2)^2 + m^2$. 
For  the massless case, we need only to set $m$ equal to zero.
Then the normalization factor  $\sqrt{(p^0 +m)/(2 p^0)}$ simplifies to $1/\sqrt{2}$.  
The first state appears with the weight $1$
and the second with the weight $- (p^1 + i p^2)/p^0$. It is evident that both solutions, 
for the massless and  massive cases are linear combinations of exactly the same number of basic states.

(Readers can easily find the second solution of Eq.(\ref{dirac3}) with the same normalization 
factor as in Eq.(\ref{dirac3sol}) but with the
factor $(p^1 -ip^2)/(|p^0| + m)$ in front of the operator 
($\gamma^1 + i \gamma^2$) and the factor one in front of the operator $(1-i\gamma^1 \gamma^2)$.)

\subsection{Solution of the Dirac equation for  spinors in four-dimensional space}

We shall  first look for one of the two solutions of the massless Weyl equation 
\begin{eqnarray}
(\gamma^a p_a =0)|\psi>
\label{weyl4}
\end{eqnarray}
treating the first irreducible representation of only left handedness from Table II.
We find
\begin{eqnarray}
|\psi> = {\cal N}\;\{\frac{1}{2}(\gamma^0 + \gamma^3)(\gamma^1 + i \gamma^2) +
\frac{p^1 + i p^2}{|p^0| + p^3} \frac{1}{2}(1 + \gamma^0 \gamma^3)
 (1-i \gamma^1  \gamma^2)\}|\psi_0> e^{-ip^a x_a},
\label{Weyl4sol}
\end{eqnarray}
or graphically
\begin{eqnarray}
|\psi> = {\cal N}\;\{\;\stackrel{03}{(-i)} \stackrel{12} {(+)} +
\frac{p^1 + i p^2}{|p^0| + p^3} \stackrel{03} {[+i]}\stackrel{12} {[-]}
 \}|\psi_0> e^{-ip^a x_a},
\label{Weyl4solgraph}
\end{eqnarray}
with ${\cal N} = \sqrt{(|p^0| +p^3)/(2 p^0)}$ and $(p^0)^2 = (p^1)^2 + (p^2)^2 + (p^3)^2. $ 

To solve a massive case 
\begin{eqnarray}
(\gamma^a p_a =m)|\psi>
\label{dirac4}
\end{eqnarray}
left- and right-handed irreducible representations are needed
\begin{eqnarray}
|\psi> &=& {\cal N}\{(|p^0| + p^3 )\frac{1}{2}(\gamma^0 + \gamma^3)
(\gamma^1 + i \gamma^2) 
+ (p^1 + i p^2) \frac{1}{2}(1 + \gamma^0  \gamma^3)
 (1-i \gamma^1  \gamma^2)  
 \nonumber\\
&  & \quad { } + m \frac{1}{2} (1 + 
 \gamma^0 \gamma^3 )( \gamma^1 + i \gamma^2)\}|\psi_0> e^{-ip^a x_a},
\label{Weyl4solmass}
\end{eqnarray}
or graphically
\begin{eqnarray}
|\psi> = {\cal N} \; \{(|p^0| + p^3 )\; \stackrel{03} {(-i)} \stackrel{12} {(+)}
+ (p^1 + i p^2) \;\stackrel{03}{[+i]} \stackrel{12} {[-]} + 
 m \stackrel{03} {[+i]} \stackrel{12} {(+)} \}|\psi_0> e^{-ip^a x_a},
\label{Weyl4solmassgraph}
\end{eqnarray}
with ${\cal N} = \sqrt{1/(2 p^0 (|p^0| + p^3))}$ and $(p^0)^2 = (p^1)^2 + (p^2)^2 + (p^3)^2 + m^2.$

\section{conclusion}
\label{conclusion}

In this paper we demonstrated  a simple technique for finding  the irreducible representations
of the Lorentz group for spinors for any-dimensional space, even or odd, of any signature, solely  
in terms of the operators 
$\gamma^a$'s, which fulfil the Clifford algebra: $\gamma^a \gamma^b + \gamma^b \gamma^a = 2\eta^{ab}$,
for which we {\em don't need to know the representation}.

By formally playing the game, that all $2^d$ linear independent polynomials of $\gamma^a$'s, 
if applied to appropriately chosen ``vacuum state'', generate $2^d$ orthogonal states, we get
$2^{d/2}$ ``families'' of twice $2^{(d-1)/2}$ Weyl spinors for $d$ even and $2^{(d+1)/2}$ families of $2^{(d-1)/2}$
Weyl spinors for $d$ odd. If a ``vacuum state'' contains the space of only one Dirac spinors, all the polynomials, 
applied to such a state, generate of course only one Dirac spinor, that is only one ``family''. 

Taking advantage of the fact that the generators of the Lorentz group are binomials of the $\gamma^a$ operators, 
enables us to find eigenstates of any operator (say the Weyl or the Dirac operator), as well as the application of 
any operator to a given state, without making a choice of the representation of the operators $\gamma^a$.

Except for the orthonormalization procedure,  this technique follows the  
refs.(\cite{norma92,norma93,norma01,holgernormadk}).

We also present a transparent graphic representations of basic states, as well as the application of 
operators to these states.

\section{Acknowledgement }

This work was supported by Ministry of Education, 
Science and Sport of Slovenia  as well as by funds CHRX -
CT - 94 - 0621, INTAS 93 - 3316, INTAS - RFBR 95
- 0567, SCI-0430-C (TSTS) of  Denmark.


\begin{thebibliography}{99}
%
\bibitem{standardmodel} T. P. Cheng and  L. F. Li, {\it Gauge Theories of Elementary Particle Physics }
(Oxford University Press, Oxford, 1986).
%
\bibitem{norma92} N. S. Manko\v c Bor\v stnik,
Spin connection as a superpartner of a vielbein,
{\it Phys. Lett.} {\bf B 292}(1992) 25-29.
%
\bibitem{norma93} N. S. Manko\v c Bor\v stnik, 
Spinor and vector representations in four dimensional
Grassmann space. {\it J. Math. Phys.} {\bf 34}, (1993)3731-3745.
%
\bibitem{normaixtapa01} N. S. Manko\v c Bor\v stnik,
Unification of spins and charges,
{\it Int. J. Theor. Phys.} {\bf 40} (2001) 315-337 and references thereby.
%
\bibitem{holgernormadk} N. S. Manko\v c Bor\v stnik and H. B. Nielsen, 
``Why odd-space and odd-time dimensions in even-dimensional spaces?'',
Phys. Lett. {\bf B 468}(2000)314-321.
%
\bibitem{kahler64} E. K\" ahler, Rend. Mat. Ser. {V \bf 21}, 452 
(1962).
%
\bibitem{norma94} N. S. Manko\v c Bor\v stnik (1994)
``Spinors, Vectors and Scalars in Grassmann Space and
Canonical Quantization for Fermions and Bosons'',
Int. Jour. Mod. Phys. {\bf A  9} 1731-1745;
``Unification of Spins and Charges in Grassmann Space'',
{\it hep-th/9408002};
``Quantum Mechanics in Grassmann Space, Supersymmetry and
Gravity'',
{\it hep-th/9406083}.
%
\bibitem{bojannorma} B. Gornik and N. S. Manko\v c Bor\v stnik, 
``Linear equations of motion for massless particles of any spin in any
even-dimensional spaces'' hep-th/0102067, hep-th/ 0102008.
%
\bibitem{norma99} Norma Susana Manko\v c Bor\v stnik (1999)
``Unification of Spins and Charges in Grassmann Space'',
{\it hep-ph/9905357};
{\it Proceedings to the International Workshop ``What Comes
Beyond the Standard Model'', Bled,
Slovenia, 29 June-9 July 1998}, Ed. by N. Manko\v c Bor\v stnik,
H. B. Nielsen, C. Froggatt, DMFA Zalo\v zni\v stvo 1999, p. 20-29.
%
\bibitem{norma01} N. S. Manko\v c Bor\v stnik, 
``Unification of spins and charges, {\it Int. J. Theor. Phys.} {\bf 40} (2001)
315-337.
\bibitem{pikanormaproceedings} A. Bor\v stnik, N. S. Manko\v c Bor\v stnik, 
``Are Spins and Charges Unified? How Can One
Otherwise Understand Connection Between Handedness (Spin) and
Weak Charge?'',
{\it Proceedings to the International Workshop on
``What Comes Beyond the Standard Model, Bled,
Slovenia, 29 June-9 July 1998},
Ed. by N. Manko\v c Bor\v stnik,
H. B. Nielsen, C. Froggatt, DMFA Zalo\v zni\v stvo 1999, p.52-57,
hep-ph/9905357, and a paper in preparation.
%
\end{thebibliography}
\end{document}